\documentclass[aps,showpacs,amsmath,amssymb,twocolumn,nofootinbib]{revtex4-1}

\usepackage{cancel}
\usepackage{cases}
\usepackage{subfigure}
\usepackage{graphicx}
\usepackage{dcolumn}
\usepackage{bm}
\usepackage{color}
\usepackage[colorlinks=true,pdfstartview=FitV,linkcolor=blue,citecolor=blue,urlcolor=blue]{hyperref}
\usepackage[mathlines]{lineno}
\usepackage[dvipsnames]{xcolor}
\usepackage{amsmath}
\usepackage{ytableau}

\everymath{\displaystyle}
\begin{document}

\newcommand{\beq}{\begin{equation}}
\newcommand{\eeq}{\end{equation}}
\newcommand{\beqs}{\begin{eqnarray}}
\newcommand{\eeqs}{\end{eqnarray}}

\title{The Upper Bound of the Second Higgs Boson Mass in Minimal Gauge Mediation with the Gravitino Warm Dark Matter}

\author{Gongjun Choi,$^{1}$}
\thanks{{\color{blue}gongjun.choi@gmail.com}}

\author{Tsutomu T. Yanagida,$^{1,2}$}
\thanks{{\color{blue}tsutomu.tyanagida@ipmu.jp}}

\author{Norimi Yokozaki,$^{3,4}$}
\thanks{{\color{blue}n.yokozaki@gmail.com}}

\affiliation{$^{1}$ Tsung-Dao Lee 
Institute, Shanghai Jiao Tong University, Shanghai 200240, China}

\affiliation{$^{2}$ Kavli IPMU (WPI), UTIAS, The University of Tokyo,
5-1-5 Kashiwanoha, Kashiwa, Chiba 277-8583, Japan}

\affiliation{$^{3}$ Zhejiang Institute of Modern Physics and Department of Physics, Zhejiang University, Hangzhou, Zhejiang 310027, China}

\affiliation{$^{4}$ Theory Center, IPNS, KEK, 1-1 Oho, Tsukuba, Ibaraki 305-0801, Japan}

\date{\today}

\begin{abstract}
A keV-scale gravitino arising from a minimal supersymmetric (SUSY) Standard Model (MSSM) is an interesting possibility since the small scale problems that the $\Lambda$CDM model encounters in the modern cosmology could be alleviated with the keV-scale gravitino serving as the warm dark matter (WDM). Such a light gravitino asks for a low scale supersymmetry (SUSY) breaking for which the gauge mediation (GM) is required as a consistent  SUSY-breaking mediation mechanism. In this paper, we show upper bounds of the masses of the second CP-even Higgs boson $H$ and the CP-odd Higgs boson $A$, assuming the keV-scale gravitino to be responsible for the current DM relic abundance: the upper bound on the mass of $H/A$ is found to be $\sim 4$ TeV for the gravitino mass of $\mathcal{O}(10$\,-\,$100)$ keV. Interestingly, the mass of $H/A$ can be as small as 2-3 TeV and the predicted $\tan\beta$ is as large as 55-60 for the gravitino mass of $\mathcal{O}(10)$ keV. This will be tested in the near future Large Hadron Collider (LHC) experiments.
\end{abstract}

\maketitle
\section{Introduction}  
The physical Higgs boson mass $125{\rm GeV}$ in the Standard Model (SM) has been regarded as one of theoretically most challenging problems to understand since it is subject to radiative corrections as large as heavy particle masses in a potential extension of the SM. Thereby this unnatural separation between the electroweak scale and a UV-cutoff such as GUT or Planck scale (a.k.a hierarchy problem~\cite{Weinberg:1975gm,Gildener:1976ai,Susskind:1978ms}) has stirred up a variety of theoretical imaginations as to a new physics beyond the SM. Among several ideas addressing the issue, supersymmetry (SUSY) is the most promising one because the systematic cancellation among the radiative corrections contributed by fermionic and bosonic degrees of freedom is inevitable consequence thereof~\cite{Maiani:1979,Veltman:1980mj} (see also \cite{Dimopoulos:1981au,Witten:1981nf,Dine:1981za,Kaul:1981hi}). 

As such, SUSY asks for a way to communicate its breaking to the visible sector. And as a plausible way, the minimal gauge mediation (MGM) model~\cite{Hamaguchi:2014sea} is advantageous in that it is free of the SUSY flavor changing neutral current (FCNC) and CP problem~\cite{Misiak:1997ei,Dine:1996ui,Gabbiani:1996hi}. In addition, in relatively low scale SUSY-breaking scenario inferred from gauge mediation, the less degree of fine-tuned cancellation is required among F-term and R-breaking contributions to a scalar potential in the theory to produce the vanishingly small cosmological constant.

In particular, for a supergravity model with the SUSY-breaking scale as low as $\mathcal{O}(1)-\mathcal{O}(10){\rm PeV}$, the mass of the gravitino ($m_{3/2}$) becomes $\mathcal{O}(10){\rm keV}$ which is attractive from a cosmological point of view. As a lightest supersymmetric particle (LSP) that is stable, neutral and very weakly interacting with MSSM particles, the gravitino can serve as a warm dark matter (WDM) candidate. The gravitino becoming free after SUSY particles are integrated-out in the MSSM thermal bath, the growth of the matter fluctuations at scales below its free-streaming length is expected to be suppressed. This may help us address small scale problems that the concordance $\Lambda$CDM suffers from~\cite{Bode:2000gq,Colin:2000dn,Destri:2012yn}. 

In this work, motivated by the aforesaid theoretical merits of the MGM implying a low SUSY-breaking scale and cosmological winning attributes of having keV-scale gravitino WDM, we consider the MGM model with the $\mathcal{O}(10){\rm keV}$ LSP gravitino. There are noteworthy features of the model. Above all, the MGM model is featured by the vanishing $B$-term at the messenger mass scale, which gives rise to a suppressed $B$-term and a large tan$\beta$ at the electroweak scale~\cite{Rattazzi:1996fb}. Relying on this distinct property, for the range of $m_{3/2}$ enabling the gravitino to resolve the cosmological small scale problems, the model is shown to predict the masses of the second CP-even Higgs boson and CP-odd Higgs boson below 4 TeV which can be tested in the future Large Hadron Collider (LHC) experiments. On top of this, the MGM model has a unstable vacuum with a life time greater than the age of the universe. Of course one may consider other gauge-mediation models with a stable SUSY-breaking vacuum (see for example Ref.~\cite{Izawa:1997gs}). In those cases, however, gaugino masses naturally become too small to survive against the LHC probe for $m_{3/2}\simeq10-100{\rm keV}$. (See, for example, Ref.~\cite{Grivaz:2008zz} for the LHC constraints on the gaugino masses) Hence, the MGM model makes the gauge-mediation consistent with the null observation of evidence of SUSY in the LHC to date.

The organization of this paper is what follows. In Sec.~\ref{sec:MGM}, we go through a brief review of the MGM model and discuss its interesting features. In Sec,.~\ref{sec:gravitinoWDM}, we discuss conditions for the gravitino to serve as a WDM candidate and how those are parametrized by the MGM model parameters. In Sec.~\ref{sec:higgsmass}, we discuss correlation between $\tan\beta$ at the electroweak scale and Higgs masses. Particularly we attend to an effect on prediction for Higgs mass in the MSSM in the large $\tan\beta$ limit. In Sec.~\ref{sec:Result}, we present the results of the analysis probing the parameter space of the model. We will make explicit discussion for how the assumption for the gravitino WDM is connected to the prediction for relatively small masses of the second CP-even Higgs boson and CP-odd Higgs boson in the MSSM. Furthermore, we discuss possible observable nature of the additional 
Higgs bosons at LHC, providing their production cross sections and decay 
channels. Lastly in Sec.~\ref{sec:conclusion}, we conclude this paper by presenting a compact summary of our assumptions, the noticeable structure of the MGM model and phenomenological consequences.


\section{Minimal Gauge Mediation Model}
\label{sec:MGM} 
Assuming a spontaneous SUSY-breaking mechanism (e.g., O’Raifeartaigh model~\cite{ORaifeartaigh:1975nky} or dynamical SUSY-breaking model with a vector-like asymptotically free gauge theory~\cite{Izawa:1996pk,Intriligator:1996pu}), for simplicity, we consider the following SUSY-breaking sector with a single chiral superfield $Z$ of which vacuum expectation value (VEV) induces the spontaneous SUSY-breaking
\beqs
\mathcal{L}_{\cancel{SUSY}}&=&\int d^{4}\theta\left[Z^{\dagger}Z-\frac{(Z^{\dagger}Z)^{2}}{4\kappa^{2}}\right]\cr\cr
&+&\int d^{2}\theta\left[-\mu_{Z}^{2}Z\right]+h.c.\,,
\label{eq:SUSYSB}
\eeqs
where $\kappa$ and $\mu_{Z}$ are dimensionful parameters given by a dynamics of the SUSY-breaking sector. Based on Eq.~(\ref{eq:SUSYSB}), the SUSY-breaking VEV of $Z$ is given by
\beq
\langle Z\rangle=0\quad,\quad\langle F_{Z}\rangle=\mu_{Z}^{2}\,.
\label{eq:VEV}
\eeq
For communicating SUSY-breaking to the MSSM sector, we introduce $N_{\rm mess}$ pairs of messenger fields ($\Psi,\overline{\Psi}$) transforming as a fundamental and an anti-fundamental representation of $SU(5)_{\rm GUT}$. The messengers of the mass $M_{\rm mess}$ are coupled to the SUSY-breaking field $Z$ via
\beq
W\supset(kZ+M_{\rm mess})\overline{\Psi}\Psi\,.
\label{eq:messenger}
\eeq
Here we assumed a $SU(5)$-invariant mass $M_{\rm mess}$ for messengers.

Note that aside from the VEV given in Eq.~(\ref{eq:VEV}), there exists another VEV which respects SUSY, i.e. $\langle Z\rangle=M_{\rm mess}/k$ and $\langle \Psi\overline{\Psi}\rangle=\mu_{Z}^{2}/k$. Thus, the vacuum with the VEV in Eq.~(\ref{eq:VEV}) is meta-stable. Nonetheless, since we assume $M_{\rm mess}^{2}>\!\!>k^{2}\mu_{Z}^{2}$, the stability of the SUSY-breaking vacuum is guaranteed for the time length longer than the age of the current universe~\cite{Hisano:2008sy}.

The gaugino and the soft SUSY-breaking scalar masses are given by
\beq
M_{a}\simeq N_{\rm mess}\frac{\alpha_{a}}{4\pi}\frac{k\mu_{Z}^{2}}{M_{\rm mess}}\,,
\label{eq:gauginomass}
\eeq
\beq
m_{\rm scalar}^{2}\simeq\sum_{a=1}^{3}2N_{\rm mess}C_{2,a}\left|\frac{\alpha_{a}}{4\pi}\frac{k\mu_{Z}^{2}}{M_{\rm mess}}\right|^{2}\,,
\label{eq:softmass}
\eeq
where $a$ specifies a SM gauge group, $\alpha_{a}\equiv g_{a}^{2}/(4\pi)$ is defined, and $C_{2,a}$ is a quadratic Casimir of a SM gauge group. $N_{\rm mess}$ denotes a number of messenger pairs and the MGM model is characterized by $N_{\rm mess}=1$. 

With all the couplings that hidden sector fields ($Z$,$\Psi$ and $\overline{\Psi}$) enjoy specified above, we assume the Higgsino mass term ($\mu$-term) is just given as the symmetry-respecting marginal operator\footnote{In this work, we do not address the possible origin of the Higgsino mass term.}
\beq
W\supset\mu H_{u}H_{d}\,,
\label{eq:muterm}
\eeq
where $H_{u}$ and $H_{d}$ are chiral superfields for the up-type and the down-type Higgs respectively. This assumption makes the model featured by the suppressed $B$-term at the scale of $M_{\rm mess}$. This is because $Z$-dependent terms of the wave function
renormalization constants of the chiral superfields $H_{u}$ and $H_{d}$ vanish
at the one-loop level. Note that MSSM scalar trilinear coupling terms ($A$-terms) are also suppressed due to the same reason. Thanks to this suppression of $A$-terms and the fact that soft scalar masses in Eq.~(\ref{eq:softmass}) are diagonal, the MGM model can avoid the FCNC problem. As can be seen later, combined with the electroweak symmetry breaking (EWSB) conditions, the resultant $B\simeq0$ at a messenger mass scale results in a somewhat large tan$\beta$ of order $\mathcal{O}(50)$, which is basically the essential point allowing for relatively small masses of the second CP-even Higgs and CP-odd Higgs.

Regarding the potential CP problem, the complex phases of gaugino mass, Higgsino mass parameter, dimensionful parameters in $A$-term and $B$-term are relevant. Above all, phases of gauginos can be rotated away by applying $U(1)_{\rm R}$. Meanwhile, Higgsino mass can be rendered real by PQ-like chiral rotation. Finally there is no concern for complex phases of parameters in $A$-term and $B$-term since those are zero at the scale of $M_{\rm mess}$ and induced by gaugino loops. Hence, the MGM model becomes free of the potential CP problem.

We end this section by clarifying the free parameters in the MGM model. For the energy scale above a messenger mass $M_{\rm mess}$, the model is described by the parameters $\mu_{Z}^{2}$ in Eq.~(\ref{eq:SUSYSB}), and $k$ and $M_{\rm mess}$ in Eq.~(\ref{eq:messenger}). After integrating out the messengers, the gaugino masses in Eq.~(\ref{eq:gauginomass}) and the soft SUSY-breaking scalar masses in Eq.~(\ref{eq:softmass}) are generated. For the energy scale below $M_{\rm mess}$, we take 
\beq
\Lambda\equiv\frac{k\mu_{Z}^{2}}{M_{\rm mess}}\quad,\quad M_{\rm mess}\,,
\label{eq:parameter}
\eeq
as a set of free parameters for the model. The later discussion in Sec.~\ref{sec:higgsmass} as to the gravitino cosmology and the second Higgs mass is to be done in the plane of ($M_{\rm mess},\Lambda$).


\section{Gravitino Warm Dark Matter}
\label{sec:gravitinoWDM} 
Along with the cosmological constant, the cold collisionless dark matter (CDM) has been successful in accounting for the large scale structure of the universe and its time evolution. This paradigm known as $\Lambda$CDM is, however, being challenged by discrepancy between its prediction for the small scale physics ($\lesssim1{\rm Mpc}$) and observation for subhalos and dwarf galaxies (e.g. core/cusp problem \cite{Moore:1999gc}, missing satellite problem \cite{Moore:1999nt,Kim:2017iwr}, too-big-to-fail problem \cite{Boylan_Kolchin_2011}). 

One of ways to alleviate the discrepancy is to consider a dark matter (DM) candidate which is relativistic at the time of its decoupling. In the presence of such a DM referred to as the WDM, the growth of the matter density fluctuation is suppressed at the scales smaller than a free-streaming length of the WDM. Interestingly, any low scale supergravity models incorporate a light gravitino which is a good candidate for the WDM. Imagining a supersymmetric universe and giving our special attention to the advantage of the gravitino WDM, in our work we consider the MSSM model with a low enough SUSY-breaking scale rendering the gravitino serve as the LSP and have a small enough mass to be the WDM candidate. 

In the attempt to constrain the gravitino mass of our interest, we borrow the definition of WDM from Ref.~\cite{Merle:2015oja} whereby the gravitino WDM is characterized by the free-streaming length lying the range of $0.01{\rm Mpc}\lesssim\lambda_{\rm FS}\lesssim0.1{\rm Mpc}$. The gravitinos are expected to decouple from the MSSM thermal bath once the MSSM superpartners are integrated-out by annihilating to the SM particles. Taking into account TeV mass scale for the superpartners, we estimate the scale factor at which the gravitino becomes the free particle by
\beq
a_{\rm dec}\sim\frac{a_{\rm EW}T_{\rm EW}}{\mathcal{O}(1){\rm TeV}}\sim\mathcal{O}(10^{-16})\,,
\label{eq:adec}
\eeq
where we used the scaling behavior of the MSSM thermal bath temperature, i.e. $T\sim1/a$, the temperature of the MSSM thermal bath at the electroweak scale $T_{\rm EW}\sim100{\rm GeV}$, and the corresponding scaling factor $a_{\rm EW}\simeq10^{-15}$.

By using Eq.~(\ref{eq:adec}), one can make a numerical estimate of a free-streaming length of the gravitino with a mass $m_{3/2}$ by following integral
\beqs
\lambda_{\rm FS}&=&\int_{t_{\rm FS}}^{t_{0}}\frac{<\!\!v_{3/2}(t)\!\!>}{a}{\rm d}t\simeq\int_{t_{\Psi}}^{t_{0}}\frac{<\!\!v_{3/2}(t)\!\!>}{a}{\rm d}t\cr\cr
&=&\int_{a_{\Psi}}^{1}\frac{{\rm d}a}{\mathcal{H}_{0}F(a)}\frac{<\!\!p_{3/2}(a_{\rm dec})\!\!>a_{\rm dec}}{\sqrt{(<\!\!p_{3/2}(a_{\rm dec})\!\!>a_{\rm dec})^{2}+m_{3/2}^{2}a^{2}}}\,, \nonumber \\
\label{eq:FSL}
\eeqs
where $<\!\!v_{3/2}(t)\!\!>$ is the averge velocity of the gravitino, $t_{\Psi}$ ($a_{\Psi}$) is the time (scale factor) when the messengers decay, $F(a)\equiv\sqrt{\Omega_{\rm rad,0}+a\Omega_{\rm m,0}+a^{4}\Omega_{\Lambda,0}}$ and $\mathcal{H}_{0}$ is the current Hubble expansion rate. Note that since $\lambda_{\rm FS}$ is dominated by the late time contribution, the two integrals in the first line of Eq.~(\ref{eq:FSL}) are almost equal to each other. At the time of decoupling, the average energy of the highly relativistic keV-scale gravitino is almost identical to its average momentum $<\!\!p_{3/2}(a_{\rm dec})\!\!>$. Thus from the ratio of the energy density to the number density of the thermal gravitino, we obtain $<\!\!p_{3/2}(a_{\rm dec})\!\!>\sim3.15T_{\rm dec}$ with $T_{\rm dec}$ the MSSM thermal bath temperature at the decoupling.

Applying the criterion $0.01{\rm Mpc}\lesssim\lambda_{\rm FS}\lesssim0.1{\rm Mpc}$, we obtain the range $10{\rm keV}\lesssim m_{3/2}\lesssim100{\rm keV}$ making the gravitino capable of addressing the small scale problem as the WDM.\footnote{For our purpose in this paper, it suffices to discuss $10{\rm keV}$ as the lower bound of $m_{3/2}$ without referring to the most recent mass constraint on the thermal WDM from the Lyman-$\alpha$ forest observation. } Given the gravitino mass $m_{3/2}=\langle F_{Z}\rangle/\sqrt{3}M_{P}$ in terms of a SUSY-breaking scale and the reduced Planck mass $M_{\rm P}\simeq2.4\times10^{18}{\rm GeV}$, this range of $m_{3/2}$ is converted into the following range of the SUSY-breaking scale
\beq
\frac{\sqrt{3}M_{\rm P}}{10^{5}}{\rm GeV}\lesssim\langle F_{Z}\rangle\lesssim\frac{\sqrt{3}M_{\rm P}}{10^{4}}{\rm GeV}\,,
\label{eq:FZ}
\eeq
where $\langle F_{Z}\rangle$ was defined in Eq.~(\ref{eq:VEV}).

We notice that avoiding too much relic abundance of the thermal gravitino of the mass $m_{3/2}\lesssim100{\rm keV}$ leads on to the severe constraint on the reheating temperature, i.e. $T_{\rm RH}\lesssim10^{3}{\rm GeV}$~\cite{Fujii:2002fv}. Bearing in mind that the lowest possible reheating temperature consistent with leptogenesis (non-thermal one) is $\sim10^{6}{\rm GeV}$~\cite{Fukugita:1986hr,Buchmuller:2005eh}, we realize that there must be a mechanism to dilute the relic abundance of the keV-scale gravitino WDM. In compliance with this reasoning, as was suggested in Ref.~\cite{Fujii:2002fv}, one may consider the possibility where a late time entropy production is made by the decay of messenger particles embedded in gauge-mediated SUSY-breaking models.  

For inducing the messenger decay to a pair MSSM particles, we introduce the following term that mixes up the messenger and MSSM supermultiplets via the R-symmetry breaking constant term in the superpotential~\cite{Fujii:2002fv}
\beq
W\supset f_{i}\frac{\langle W\rangle}{M_{\rm P}^{2}}\Psi\overline{5}_{i}=f_{i}m_{3/2}\Psi\overline{5}_{i}\,,
\label{eq:messengerdecay}
\eeq
where the subscript $i$ is the generation index and $f_{i}$ is a dimensionless coefficient. As the consequence of the mixing in Eq.~(\ref{eq:messengerdecay}), an operator for the messenger decay to $H_{d}$ and $10_{i}$ is induced with the coupling proportional to $f_{i}m_{3/2}/M_{\rm mess}$. Thus the lightest scalar component of the messenger weak doublet can decay to the higgsino and the SM lepton with the decay rate
\beq
\Gamma_{\Psi}\simeq\frac{1}{8\pi}\left(\frac{m_{\tau}}{v\cos\beta}\right)^{2}\left(\frac{f_{3}m_{3/2}}{M_{\rm mess}}\right)^{2}M_{\rm mess}\,,
\label{eq:GammaPsi}
\eeq
where $v\simeq\sqrt{v_{u}^{2}+v_{d}^{2}}\simeq174{\rm GeV}$ is assumed with $v_{u}$ ($v_{d}$) the up (down)-type Higgs VEV. We note that it is demanded for the model to have large enough $f_{i}$s so as to complete $\Psi$-decay before the big bang nucleosynthesis (BBN) era begins. Otherwise, the primordial light elements formed during BBN time can be destroyed by the electrically charged high energy decay products of $\Psi$-decay and inconsistent deficit of the primordial light elements is caused.

By comparing the decay rate of the messenger in Eq.~(\ref{eq:GammaPsi}) to the Hubble expansion rate during the radiation-dominated era, i.e. $\Gamma_{\Psi}\sim \mathcal{H}\simeq T^{2}/M_{\rm P}$, we can obtain the MSSM thermal bath temperature $T_{\rm \Psi}$ at the time when the messenger decays. $T_{\Psi}$ is given by~\cite{Fujii:2002fv}
\beqs
T_{\Psi}&\simeq&33{\rm MeV}\times\frac{f_{i}}{k}\left(\frac{10}{g_{*}(T_{\Psi})}\right)^{1/4}\cr\cr&\times&\left(\frac{\Lambda}{10^{5}{\rm GeV}}\right)\times\left(\frac{M_{\rm mess}}{10^{8}{\rm GeV}}\right)^{1/2}\,.
\label{eq:TPsi}
\eeqs

Now the ratio of the entropy density of radiation coming from the decay of the messenger to that of existing MSSM radiation is given by $\Delta\equiv s_{\Psi}/s_{\rm rad}=(4/3)(M_{\rm mess}Y_{\rm mess}/T_{\Psi})$ where $Y_{\rm mess}=3.65\times10^{-10}(M_{\rm mess}/10^{6}{\rm GeV})$ is the comoving number density of the messenger~\cite{Dimopoulos:1996gy}. For a large enough $\Delta$, the entropy of the universe is dominated by that of the messenger, which makes it possible to dilute overproduction of the thermal gravitino. For the gravitino to be WDM today, parameters controlling $\Delta$ should satisfy
\beqs
\Omega_{\rm DM}h^{2}&=&\frac{\Omega_{\rm 3/2}h^{2}}{\Delta}\cr\cr
&\simeq&0.16\left(\frac{f_{3}}{0.1}\right)\left(\frac{g_{*,{\rm MSSM}}}{10}\right)^{\frac{-1}{4}}\left(\frac{\tan\beta}{50}\right)\cr\cr&\times&\left(\frac{m_{3/2}}{10{\rm keV}}\right)^{2}\left(\frac{10^{8}{\rm GeV}}{M_{\rm mess}}\right)^{\frac{5}{2}}\,,
\label{eq:gravitinoabundance}
\eeqs
where $\Omega_{3/2}$ is the relic abundance for the thermal gravitino, $h$ is defined via $\mathcal{H}_{0}=100h({\rm km}/{\rm Mpc}/{\rm sec})$ and $g_{*,{\rm MSSM}}=228.75$ is the effective degrees of freedom of MSSM particles after the messengers decay. In Sec.~\ref{sec:Result}, we shall discuss whether Eq.~(\ref{eq:gravitinoabundance}) can be satisfied by $(\tan\beta,M_{\rm mess})$ of our interest for $m_{3/2}\simeq10-100{\rm keV}$.


\section{Higgs Masses in the MGM model}
\label{sec:higgsmass}
As was pointed out in Sec.~\ref{sec:MGM}, $B$-term at the scale of $M_{\rm mess}$ vanishes due to the set-up of the MGM model (see Eq.~(\ref{eq:muterm}) and the associated text), and the B-term at a low energy scale is generated  by gaugino masses through radiative corrections. 
The renormalization group equation (RGE) of the parameter $B$ consists of contributions proportional to gaugino masses and scalar trilinear couplings, where the scalar trilinear couplings are also radiatively generated from the gaugino masses. Since these two contributions are loop suppressed and have opposite signs,\footnote{When the $\mu$-term has the positive sign, the $B$-term should be positive at the low energy scale, which is achieved if the contributions proportional to the scalar trilinear couplings are larger than the contributions proportional to the gaugino masses. In contrast, for the negative $\mu$ case, the $B$-term should be negative at the low energy scale, which is achieved if the contributions proportional to the gaugino masses are larger.}
the MGM model tends to produce a smaller $B$ value at the low energy scale than other models.

Together with soft masses for $H_{u}$ and $H_{d}$, the parameter $B$ obtained at the electroweak scale via RGE can determine values of $\tan\beta$ and $\mu$-parameter with the aid of the following two conditions for the EWSB ($\partial V/\partial H_{u}^{0}=\partial V/\partial H_{d}^{0}=0$)
\beq
\frac{m_{Z}^{2}}{2}=\frac{(m_{H_{d}}^{2}+\frac{1}{2v_{d}}\frac{\partial(\Delta V)}{\partial v_{d}})-(m_{H_{u}}^{2}+\frac{1}{2v_{u}}\frac{\partial(\Delta V)}{\partial v_{u}})\tan^{2}\beta}{\tan^{2}\beta-1}-\mu^{2}\,,
\label{eq:EWSB1}
\eeq
\beqs
&&B\mu(\tan\beta+\cot\beta)\cr\cr
&=&m_{H_{u}}^{2}+\frac{1}{2v_{u}}\frac{\partial(\Delta V)}{\partial v_{u}}+m_{H_{d}}^{2}+\frac{1}{2v_{d}}\frac{\partial(\Delta V)}{\partial v_{d}}+2\mu^{2}\,,\nonumber \\
\label{eq:EWSB2}
\eeqs
where $m_{Z}$ is the $Z$-boson mass, $m_{H_{u}}$ ($m_{H_{d}}$) is the soft mass for $H_{u}$ ($H_{d}$) and $\Delta V$ is a radiative correction to the Higgs potential. Note that the left hand side of Eq.~(\ref{eq:EWSB2}) is nothing but CP-odd Higgs mass squared, i.e. $m_{A}^{2}=B\mu(\tan\beta+\cot\beta)$.

As is well known, the RGE for $m_{H_{u}}^{2}$ is subject to the negative contributions attributable to Yukawa couplings, which gives rise to the negative sign of $m_{H_{u}}^{2}$ at the electroweak scale. For a large $\tan\beta$ case, simplification of Eq.~(\ref{eq:EWSB1}) to remove $m_{H_{d}}^{2}$ dependence and substitution of the resultant expression of $2\mu^{2}$ into Eq.~(\ref{eq:EWSB2}) yields
\beq
m_{A}^{2}\simeq m_{H_{d}}^{2}+\frac{1}{2v_{d}}\frac{\partial(\Delta V)}{\partial v_{d}}
-m_{H_{u}}^{2}-\frac{1}{2v_{u}}\frac{\partial(\Delta V)}{\partial v_{u}}\, , \ (\tan\beta>\!\!>1)
\label{eq:Higgsmassrelation}
\eeq
Now remarkably for a large enough $\tan\beta$, we notice that not only $m_{H_{u}}^{2}$ but also $m_{H_{d}}^{2}$ could be negative at the electroweak scale. This enables cancellation between the two results, allowing for the smaller $m_{A}^{2}$ value than the case with a small $\tan\beta$. Essentially this is attributed to large Yukawa couplings for the tau lepton and the bottom quark resulting from a large $\tan\beta$.

For eigenvalues of the mass matrix for the CP-even Higgs ($m_{H}$ and $m_{h}$ with $m_{H}>m_{h}$), one obtains 
\beq
m_{A}\simeq m_{H}
\eeq
when $m_{A}>\!\!>m_{Z}$ is satisfied. This predicts a relatively light second Higgs mass ($m_{H}$) comparable to a relatively light $m_{A}$ for a large $\tan\beta$. In the next section, we shall see how (1) large $\tan\beta$ could arise for the model parameter space producing keV-scale gravitino and (2) how light $m_{H}$ and $m_{A}$ could be for the resulting large $\tan\beta$.


\section{Results of Analysis}
\label{sec:Result}

\begin{figure}[htp]
  \centering
  \hspace*{-5mm}
  \includegraphics[scale=0.65]{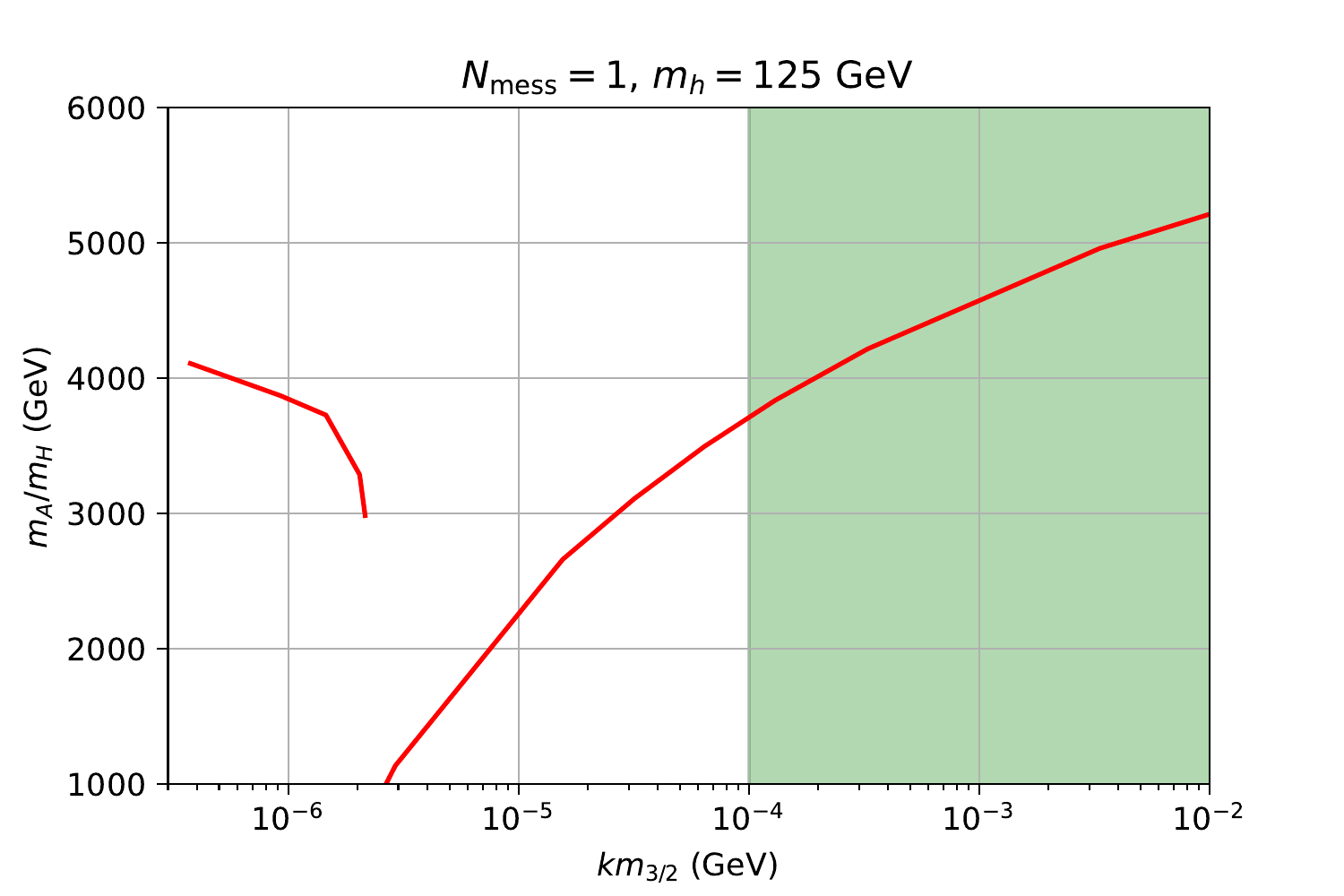}
  \caption{The plot of the second CP even Higgs mass ($m_{H}$) and the CP odd Higgs mass ($m_{A}$) as a function of $km_{3/2}$. The green shaded region is excluded when the MGM model produces the gravitino WDM.}
  \vspace*{-1.5mm}
\label{fig:1}
\end{figure}

In this section, we discuss the parameter space $(M_{\rm mess},\Lambda)$ of the MGM model yielding the consistent values of the lightest CP-even Higgs mass $(m_{h})$ and the corresponding resultant second CP-even Higgs mass ($m_{H}$). $\Lambda$ was defined in Eq.~(\ref{eq:parameter}). To this end, we perform the analysis of solving the RGE equations for MSSM parameters and computing MSSM particle mass spectra with the aid of {\tt SOFTSUSY} package~\cite{Allanach:2001kg}. Following this, the Higgs mass spectra is obtained by {\tt FeynHiggs 2.16.1}~\cite{Heinemeyer:1998yj,Heinemeyer:1998np,Degrassi:2002fi,Frank:2006yh,Hahn:2013ria,Bahl:2016brp,Bahl:2017aev,Bahl:2018qog}.

Firstly, we show in Fig.~\ref{fig:1} the second CP even Higgs mass ($m_{H}$) and CP odd Higgs mass ($m_{A}$) as a function of $km_{3/2}$ where $k$ was defined in Eq.~(\ref{eq:messenger}). The regime of $km_{3/2}$ smaller than shown in the horizontal axis is irrelevant since the messengers become tachyonic. For the two disconnected red lines, the left one corresponds to $\mu<0$ while the right one does to $\mu>0$. The left line is cut at $km_{3/2}\simeq2\times10^{-6}{\rm GeV}$ since EWSB fails to occur for the larger $km_{3/2}$. 

For $km_{3/2}\lesssim100{\rm keV}$, we observe that the second CP even Higgs mass is upper-bounded by $\sim4{\rm TeV}$ independent of the sign of $\mu$-parameter. Remarkably, this implies that the MGM model predicts for the upper bound $\sim4{\rm TeV}$ of the second CP even Higgs mass provided it has the gravitino as the WDM candidate. This is because $km_{3/2}\simeq100{\rm keV}$ coincides with the upper bound of the gravitino WDM mass for a perturbative $k\lesssim1$. On the other hand, it can be observed that $m_{H}/m_{A}$ below 3TeV is allowed in Fig.~\ref{fig:1} and this region of $m_{H}/m_{A}$ is crucial particularly for the LHC search of the second Higgs boson. Therefore, a reasonable question can be how large a parameter space the model has for $m_{H}/m_{A}\lesssim3{\rm TeV}$. Since this region corresponds to $\mu>0$, below we probe the parameter space for $\mu>0$ case.

\begin{figure*}[htp]
  \centering
  \hspace*{-5mm}
  \subfigure{\includegraphics[scale=0.65]{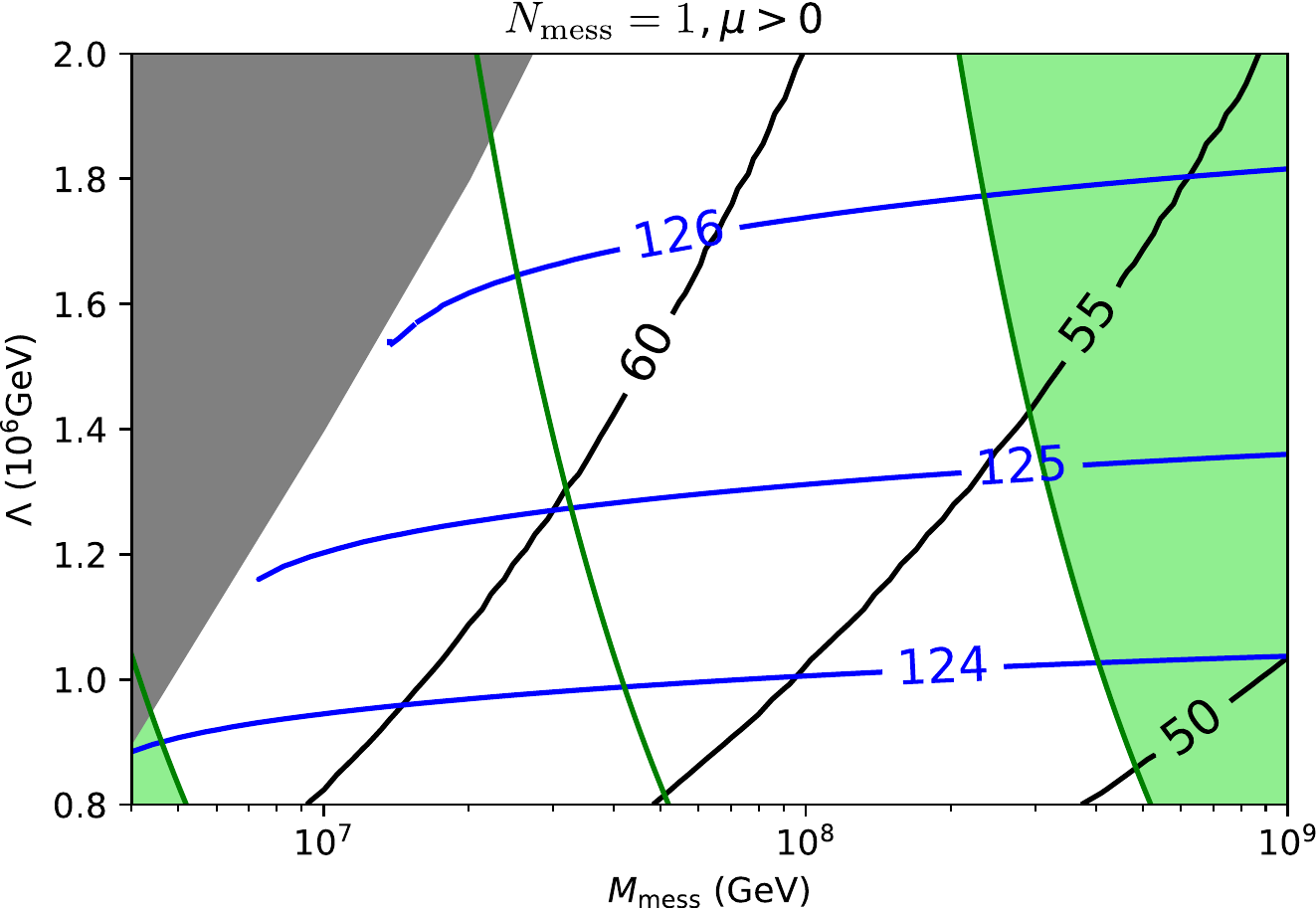}}\qquad
  \subfigure{\includegraphics[scale=0.65]{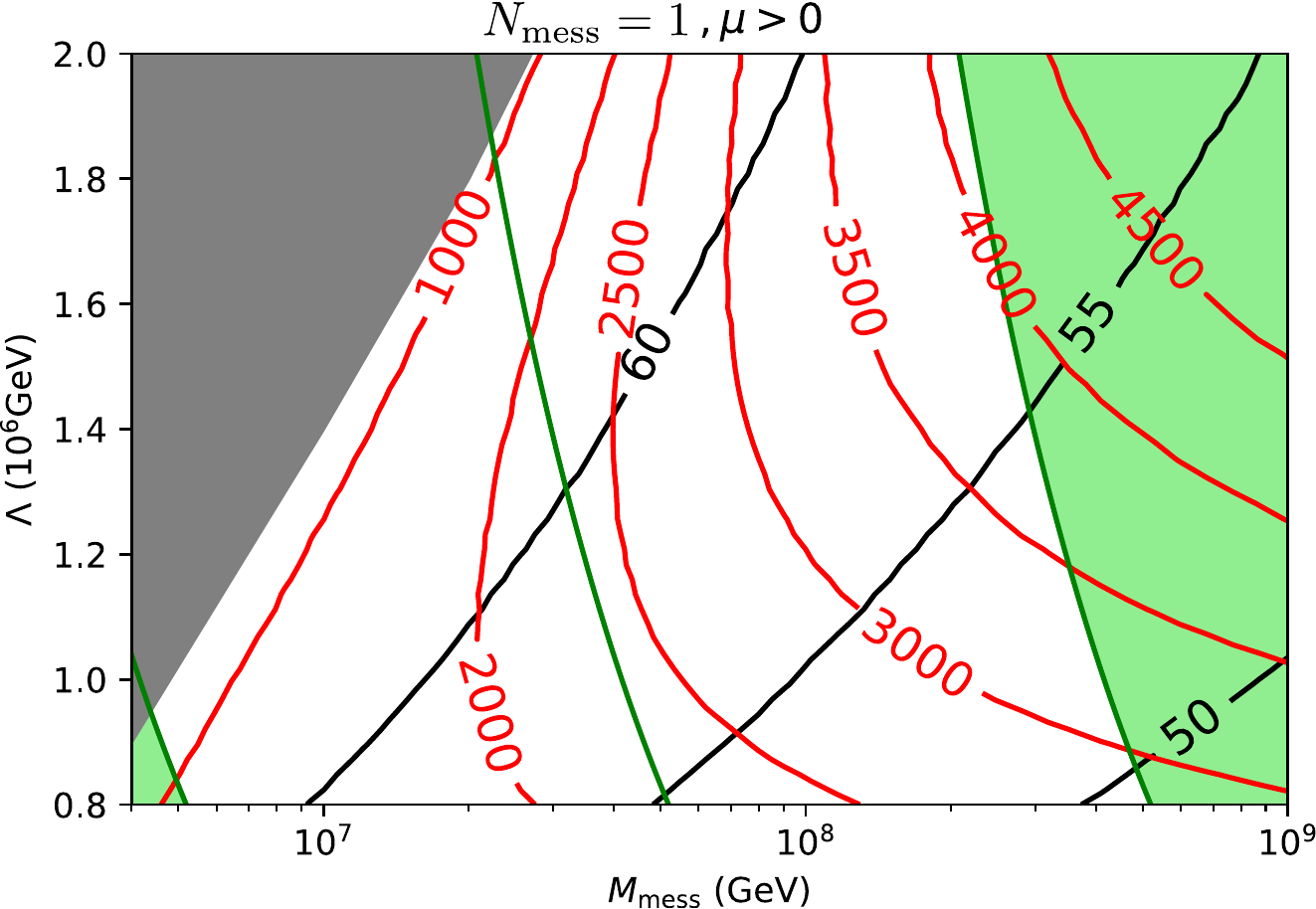}}
  \caption{Parameter space of the model for $\mu>0$ case. In the left panel, for a pair of messengers ($\Psi,\overline{\Psi}$) in ($5$,$\overline{5}$) representation of $SU(5)_{\rm GUT}$ respectively, we show values of the set of parameters ($M_{\rm mess},\Lambda$) producing the lightest CP-even Higgs mass $124{\rm GeV}$, $125{\rm GeV}$ and $126{\rm GeV}$ (blue solid line). $\Lambda\equiv k\langle F_{Z}\rangle/M_{\rm mess}$ was defined in Eq.~(\ref{eq:parameter}). In the right panels, each red line is a set of points yielding the specified second CP-even Higgs mass $m_{H}$ in the unit of GeV. For both panels, the black lines are the group of points in the parameter space corresponding to the specified $\tan\beta$ values. Each green line is the set of points giving $km_{3/2}=1,10$ and $100{\rm keV}$ from the left. For the gray shaded region, EWSB cannot happen. The left (right) green shaded region correspond to $km_{3/2}<1{\rm keV}$ ($km_{3/2}>100{\rm keV}$). }
  \vspace*{-1.5mm}
\label{fig:2}
\end{figure*}

In Fig.~\ref{fig:2}, we show Higgs boson masses in the plane of $(M_{\rm mess},\Lambda)$. In the left and right panel, using blue and red lines, we show values of $(M_{\rm mess},\Lambda)$ yielding the specified the lightest CP even Higgs mass and the second CP even Higgs mass, respectively. For both panels, each black line shows a set of points yielding the specified $\tan\beta$ value. Note that the larger $M_{\rm mess}$ enhances $B$-parameter value via logarithmic dependence arising from the RGE and thus corresponds to a smaller $\tan\beta$ for a fixed $\Lambda$. Also displayed are three green lines corresponding to $km_{3/2}=1,10,100{\rm keV}$ from the left in the Fig.~\ref{fig:2}. 

\begin{table}[t]
\centering
\caption{The $b$-associated production cross-section of $H/A$ for the MGM model with $m_{h}=125{\rm GeV}$ and $\mu>0$.}
\begin{tabular}{|c|c|} \hline
 $m_{H}$[TeV] & $\sigma[{\rm pb}]$ \\
\hline\hline
  2.2  &  0.0054  \\
  2.4  &  0.0028 \\
  2.6 &  0.0015  \\
  2.8  &  0.00080  \\
 3.0 &  0.00045   \\
  3.2  &  0.00026 \\
 3.4 &  0.00015  \\
\hline
\end{tabular}
\label{table:sigma} 
\end{table}

\begin{table}[t]
\centering
\caption{Mass spectra in the MGM model with $\mu>0$.}
\label{tab:2}
\begin{tabular}{|c||c|c|}
\hline
Parameters & Point {\bf I} & Point {\bf II} \\ 
\hline
$\Lambda$ (TeV) & 1300  & 1700  \\
$M_{\rm mess}$ (GeV) & $5.0 \times 10^7$  & $3.5 \times 10^7$  \\
\hline
\hline
Particles & Mass (TeV) & Mass (TeV) \\
\hline
$\tilde{g}$ & 8.3 &  10.6\\
$\tilde{q}$  & 10.8-11.6 & 14.0-15.1\\
$\tilde{t}_{1,2}$ & 9.6, 10.6 & 12.6, 13.7\\
$\tilde{b}_{1,2}$ & 9.8, 10.6 & 12.6, 13.7\\
$\tilde{e}_{L, R}$ & 4.6, 2.5 & 5.9, 3.2\\
$\tilde{\tau}_{1,2}$ & 1.7, 4.4 & 2.2, 5.7\\
$\tilde{\chi}_{1,2}^0$ & 1.8, 3.4 & 2.4, 4.4\\
$\mu$ & 4.3 & 5.2\\
$\tilde{\chi}^{\pm}_{1}$ & 3.4 & 4.4\\
$h_{\rm SM\mathchar`-like}$ (GeV) & 125 & 126\\
$H/A$ (GeV) & 2660 & 2220 \\
\hline
$\tan\beta$& 60.1 & 63.1 \\
$k m_{3/2}$ (keV) & 15.6   &  14.3   \\
\hline
\end{tabular}
\label{table:mass} 
\end{table}

\begin{figure*}[htp]
  \centering
  \hspace*{-5mm}
  \subfigure{\includegraphics[scale=0.65]{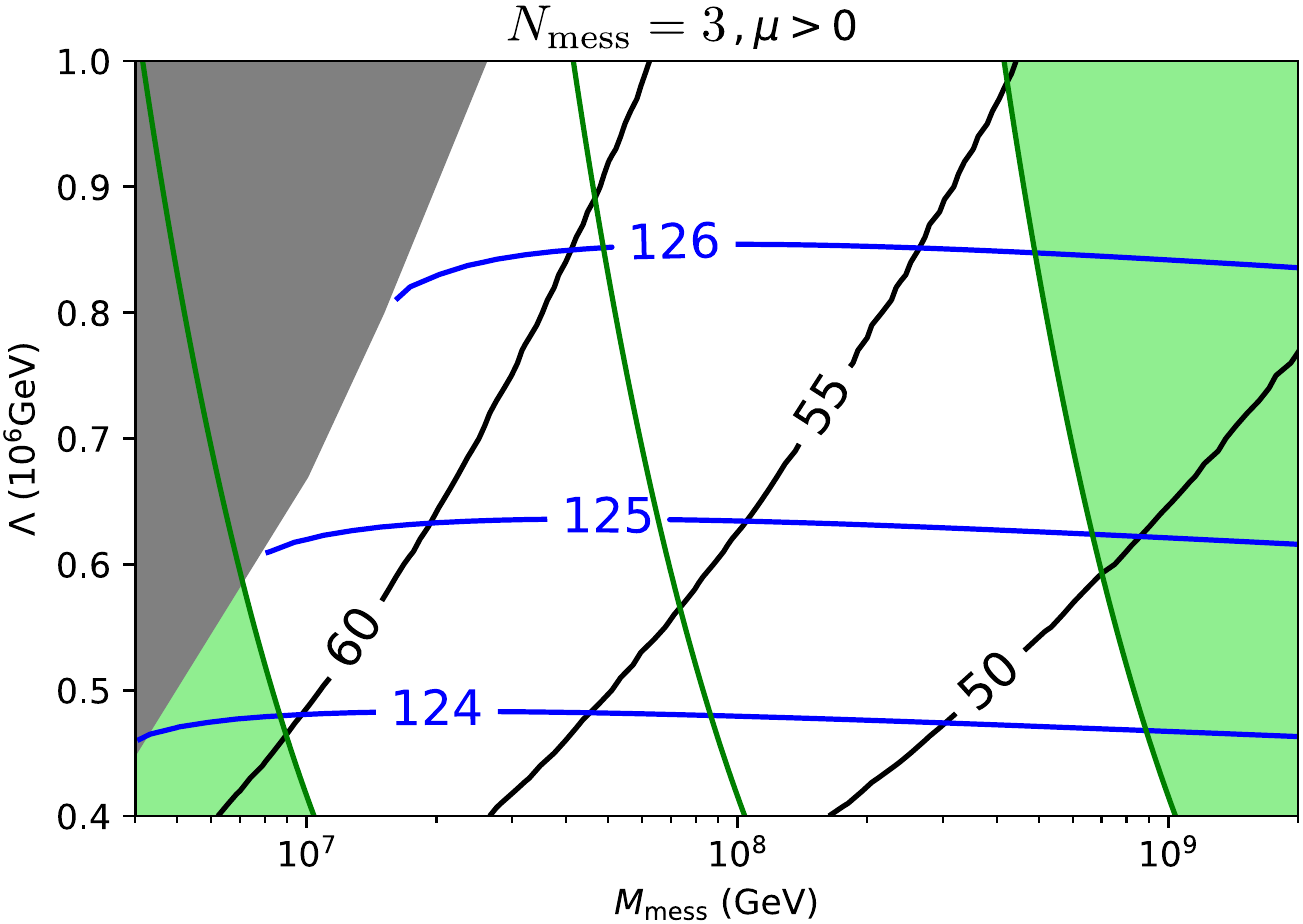}}\qquad
  \subfigure{\includegraphics[scale=0.65]{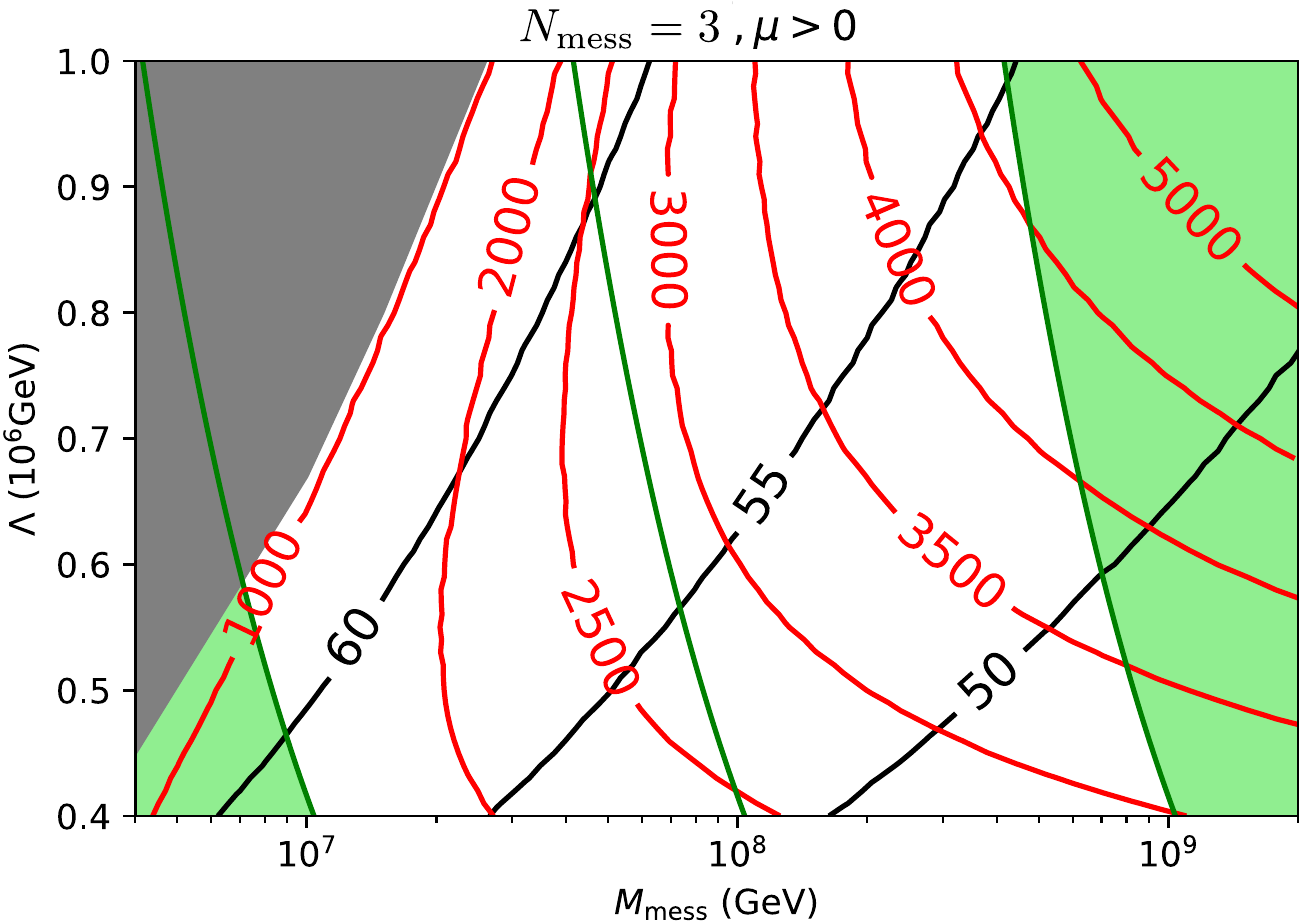}}
  \caption{Parameter space of the model for $\mu>0$ case. In the left panel, for three pairs of messengers ($\Psi,\overline{\Psi}$) in ($5$,$\overline{5}$) representation of $SU(5)_{\rm GUT}$ respectively, we show values of the set of parameters ($M_{\rm mess},\Lambda$) producing the lightest CP-even Higgs mass $124{\rm GeV}$, $125{\rm GeV}$ and $126{\rm GeV}$ (blue solid line). $\Lambda\equiv k\langle F_{Z}\rangle/M_{\rm mess}$ was defined in Eq.~(\ref{eq:parameter}). In the right panels, each red line is a set of points yielding the specified second CP-even Higgs mass $m_{H}$ in the unit of GeV. For both panels, the black lines are the group of points in the parameter space corresponding to the specified $\tan\beta$ values. Each green line is the set of points giving $km_{3/2}=1,10$ and $100{\rm keV}$ from the left. For the gray shaded region, EWSB cannot happen. The left (right) green shaded region correspond to $km_{3/2}<1{\rm keV}$ ($km_{3/2}>100{\rm keV}$). }
  \vspace*{-1.5mm}
\label{fig:3}
\end{figure*}

Assuming $k=\mathcal{O}(0.1)-\mathcal{O}(1)$, one can see that $10{\rm keV}\lesssim m_{3/2}\lesssim100{\rm keV}$ is consistent with the region between the rightmost ($km_{3/2}=100{\rm keV}$) and the leftmost ($km_{3/2}=1{\rm keV}$) green lines. The right green shaded region gives $km_{3/2}>100{\rm keV}$. Thus, it is excluded since $m_{3/2}>100{\rm keV}$ holds there for a perturbative $k\lesssim1$. In other words, the MGM model cannot have the gravitino WDM resolving the small scale problems there. Both the blue line for $125{\rm GeV}$ and the right green shaded region being taken into account together simultaneously, we realize that the upper bound of $m_{H}$ lies in $3.5-4{\rm TeV}$. By referring to the excluded region in $(m_{A},\tan\beta)$ plane ($95\%$ C.L.) given in Ref.~\cite{Aad:2020zxo}, we notice that $m_{H}\lesssim2{\rm TeV}$ (equivalently $m_{A}\lesssim2{\rm TeV}$) and $\tan\beta>60$ are excluded due to the large production cross section (times branching ratio). For the gray shaded region, EWSB does not occur. In order to demonstrate that a non-minimal model has an interesting parameter space as large as the minimal case,  in Fig.~\ref{fig:3}, we show the result of analysis for the case with $N_{\rm mess}=3$. One can see that almost similar interesting parameter space arises even for $N_{\rm mess}=3$ case.


Finally, we notice that points shown in Fig.~\ref{fig:2} are accompanied by $\tan\beta\simeq50-60$ and $m_{3/2}\simeq10-100{\rm keV}$. Referring to Eq.~(\ref{eq:gravitinoabundance}), we realize that the correct relic abundance matching between the gravitino and the current DM population requires $f_{3}=\mathcal{O}(10^{-2})-\mathcal{O}(10^{-1})$. Note that $\Lambda$ remains almost constant in the viable parameter space in Fig.~\ref{fig:2} and thus roughly $\Omega_{3/2}h^{2}/\Delta$ becomes proportional to $M_{\rm mess}^{-1/2}$. Hence, even if $M_{\rm mess}$ changes by two orders of magnitude in Fig.~\ref{fig:2} in the viable parameter space, $f_{3}$ needs to change only by one order of magnitude for DM relic density matching. We checked that this $f_{3}$ is large enough to induce the decay of the messenger particle to the Higgsino and the SM lepton before the BBN era is reached.\footnote{From Eq.~(\ref{eq:TPsi}), one can see that for $\Lambda\simeq10^{6}{\rm GeV}$ and $M_{\rm mess}=\mathcal{O}(10^{7})-\mathcal{O}(10^{8}){\rm GeV}$, one can see $T_{\Psi}>1{\rm MeV}$ is satisfied indeed.} Therefore, the model can indeed produce the thermal warm gravitino dark matter for the parameter space of $(M_{\rm mess},\Lambda)$ of our interest.

We conclude this section by discussing (1) information useful for experimental searching for the second Higgs and (2) particle mass spectra for points in viable parameter space in Fig.~\ref{fig:2}. For the minimal MGM model with $m_{h}=125{\rm GeV}$ and $\mu>0$, using {\tt HDECAY}~\cite{Djouadi:1997yw}, we find that the branching ratios of the the second Higgs decay for the main decay modes read BR($H\rightarrow b+\overline{b}$)$\simeq0.78$ and BR($H\rightarrow \tau+\overline{\tau}$)$\simeq0.22$. In addition,  shown in Table.~\ref{table:sigma} is the $b$-associated production cross-section of $H/A$ for the MGM model with $m_{h}=125{\rm GeV}$ and $\mu>0$, which is obtained by using {\tt SusHi} package~\cite{Harlander:2012pb,Harlander:2016hcx}. To help reader's understanding, in Table.~\ref{tab:2}, we display the particle mass spectra of the model for two selective points in viable parameter space in Fig.~\ref{fig:2}. The point {\bf I} ({\bf II}) lies in the blue line of 125GeV (126GeV) in the left panel. For both cases, one can see that stau becomes NLSP of the model. However, observation of the stau seems very challenging as can be seen in Ref.~\cite{CMS-PAS-FTR-18-010}.


\section{Conclusion}
\label{sec:conclusion}
The minimal gauge mediation (MGM) model is appealing in that it is free of FCNC and CP problems. Moreover, $\mu$-term is present in the model just as a marginal operator, which makes $B$-term vanish at a messenger mass scale. Accordingly, the model is featured by a rather small $B$-term and a large $\tan\beta$ at the electroweak scale. The resultant large $\tan\beta$, in turn, permits cancellation between $m_{H_{u}}^{2}$ and $m_{H_{d}}^{2}$  in Eq.~(\ref{eq:Higgsmassrelation}), opening up the interesting possibility to have relatively small masses for second CP-even Higgs and CP-odd Higgs.

This structure of the MGM model alone, nevertheless, does not necessarily predict a light second CP-even Higgs boson simply because there is no upper bound on a SUSY-breaking scale or a messenger mass scale. On the other hand, the small scale issues continue to challenge $\Lambda$CDM model in the modern cosmology and assuming a WDM candidate can help us resolve the issue. Notably the gravitino can play a role of the WDM provided its mass lies in $10-100{\rm keV}$ range. Inspired by this cosmological advantage that the keV-scale gravitino can enjoy, we considered the scenario within the MGM model where the keV-scale gravitino becomes the WDM today with the late time entropy production triggered by the decay of messenger particles. This assumption helped us narrow down interesting parameter space of the model.

So obtained parameter space of $(M_{\rm mess},\Lambda)$ was shown to be able to be consistent with the observed Higgs mass 125GeV when the mass of the lightest CP-even Higgs is identified with 125GeV. For values of $(M_{\rm mess},\Lambda)$ achieving the consistency, we computed values of $\tan\beta$ and the second CP-even Higgs mass $m_{H}$. To our surprise, it turns out that the upper bounds of $m_{H}$ and $m_{A}$ are as small as $\sim4{\rm TeV}$ for ($M_{\rm mess}=\mathcal{O}(10^{7})-\mathcal{O}(10^{8}){\rm GeV},\Lambda=\mathcal{O}(10^{6}){\rm GeV}$) in the MGM model. In particular, the mass of the second Higgs (and the CP-odd Higgs) is as small as 2-3 TeV for $\mathcal{O}(10)$ keV gravitino with $k=1$, and the predicted $\tan\beta$ is as large as 55-60. Due to the large $\tan\beta$, the Yukawa coupling to bottom quarks becomes $\sim 1$, generating the large production cross section for those Higgs bosons as shown in Table.~\ref{table:sigma}. Because of the above two reasons we expect observation/exclusion of the additional Higgs bosons in MSSM at the future LHC experiments at 14 TeV run, providing us with the opportunity to test supersymmetry with the warm gravitino dark matter.


\begin{acknowledgments}
N. Y. is supported by JSPS KAKENHI Grant Number JP16H06492. T. T. Y. is supported in part by the China Grant for Talent Scientific Start-Up Project and the JSPS Grant-in-Aid for Scientific Research No. 16H02176, No. 17H02878, and No. 19H05810 and by World Premier International Research Center Initiative (WPI Initiative), MEXT, Japan. 

\end{acknowledgments}


\bibliography{main}

\end{document}